\documentclass[twocolumn]{jpsj2} %% two-column layout
\usepackage{bm} %bold math

\title{Commensurate Itinerant Antiferromagnetism in BaFe$_2$As$_2$:
$^{75}$As-NMR Studies on a Self-Flux Grown Single Crystal}

\author{Kentaro~\textsc{Kitagawa}\thanks{kitag@issp.u-tokyo.ac.jp},
Naoyuki~\textsc{Katayama}, Kenya~\textsc{Ohgushi}, Makoto~\textsc{Yoshida}, and
Masashi~\textsc{Takigawa}}

\inst{Institute for Solid State Physics, University of Tokyo,
5-1-5 Kashiwanoha, Kashiwa, Chiba 277-8581, Japan}

\abst{We report results of $^{75}$As nuclear magnetic resonance (NMR) 
experiments on a self-flux grown single crystal of BaFe$_2$As$_2$. A first-order
antiferromagnetic (AF) transition near 135~K was detected by the splitting of NMR lines, 
which is accompanied by simultaneous structural transition as evidenced by a sudden large change 
of the electric field gradient tensor at the As site. The NMR results lead almost uniquely to the 
stripe spin structure in the AF phase.  The data of spin-lattice relaxation rate 
indicate development of anisotropic spin fluctuations of the stripe-type with
decreasing temperature in the paramagnetic phase.}

\kword{BaFe2As2, NMR, itinerant antiferromagnetism, ternary iron arsenide}

\begin{document}
\maketitle

\section{Introduction} 
The discovery of high temperature superconductivity in the layered oxypnictide compound 
LaFeAsO$_{1-x}$F$_x$ by Kamihara \textit{et al.}\cite{KamiharaJACS} with the 
superconducting transition temperature $T_\text{c} = 26$~K has opened new opportunities to investigate 
interplay between magnetism and superconductivity.  The unsubstituted compound LaFeAsO 
undergoes successive phase transitions: a tetragonal to orthorhombic structural
transition at 155~K\cite{Cruz1111NS} followed by an antiferromagnetic (AF) transition at
142~K\cite{Nakai1111NMR}. These transitions are suppressed by substituting
oxygen with fluorine, leading to the superconductivity.
Higher $T_\text{c}$ exceeding 50~K was achieved by substituting La with other
rare earth elements\cite{KitoNd1111,YangGd1111}.

Subsequent synthesis of a new ternary series of superconductors without oxygen,
$A_{1-x}B_x$Fe$_2$As$_2$ ($A$=Ba, Sr, Ca; $B$=K,
Na)\cite{RotterBa122SDW,RotterBK122,SasmalSrK122,WuCaNa122} has marked a further important step. For these
materials, large single crystals can be grown by flux methods,
which is crucial to investigate anisotropic properties of 
the layered superconductors and to determine the paring symmetry.  
Here again, the unsubstituted materials show 
both the structural and the AF transitions but at the same temperature.  For example, 
BaFe$_2$As$_2$ has a tetragonal structure with the space group $I4/mmm$ at 
room temperature.  The structural transition to the orthorhombic $Fmmm$ 
space group and the AF transition into a stripe-type order take place simultaneously
near 140~K\cite{RotterBa122SDW,HuangBa122NS}.  Superconductivity emerges by 
substituting Ba with K, which suppresses the structural and the AF transitions. 
Whether the structural instability or magnetic fluctuations (or both) plays
vital role for the occurrence of superconductivity is an important issue to be 
addressed by various microscopic experiments. 

In this paper, we report nuclear magnetic resonance (NMR) studies on $^{75}$As 
nuclei in a single crystal of BaFe$_2$As$_2$ grown in FeAs flux.  NMR is a powerful tool, 
sensitive to both the magnetism and the local structure.  We observed a discontinuous splitting 
of the NMR lines below 135~K, indicating a first-order AF transition.  A simultaneous 
tetragonal-to-orthorhombic structural transition was detected by a sudden
change of the quadrupole splitting in the NMR spectra, indicating substantial change in
the local charge distribution around the As sites.  Observation of the internal field parallel to 
the $c$-axis in the AF state leads nearly uniquely to the stripe-type spin
structure with the AF moment perpendicular to the stripe.  The temperature dependence of the 
spin-lattice relaxation rate (1/$T_1$) indicates development of  anisotropic 
spin fluctuations of stripe-type in the paramagnetic tetragonal phase. 

\section{Experiment}
The sample used in our NMR experiment was prepared by the self-flux method.
To avoid contamination by external elements, we chose FeAs as the flux after 
Wang~\textit{et\,al.}\cite{WangBa122SC} The starting materials were mixed in a
graphite crucible with the ratio Ba:Fe:As=1:4:4 and sealed in a double quartz tube with 
argon gas.  The tube was heated up to 1100{$^\circ$}C in 14 hours (including 
the holding at 700{$^\circ$}C for 3 hours) and slowly cooled down to
900{$^\circ$}C in 50 hours.
The resistivity and the magnetic susceptibility showed a sharp transition at $\sim 135$~K, in agreement with 
the results by Wang~\textit{et\,al.}\cite{WangBa122SC}

For the NMR experiments, a crystal with the size $3\times 2\times 0.05$~mm$^{3}$ 
was mounted on a probe equipped with a two-axis goniometer.  This allows fine 
alignment of arbitrary crystalline axis along the magnetic field within
$\sim$0.2~{$^\circ$}. The field-swept NMR spectra were taken by Fourier
transforming the spin-echo signal with the step-sum technique. The Knight shift and the spin-lattice
relaxation rate ($T_1^{-1}$) were measured at the fixed field of 6.615~T.
The value of $T_1^{-1}$ was determined by 
fitting the time dependence of the spin-echo intensity of the central transition line after 
the inversion pulse to the theoretical formula\cite{NarathTiNMR}. 
Good fitting was obtained above 20~K.
At lower temperatures, an extra component with short $T_{1}$ 
caused slight deviation, which is likely due to residual disorder.

\begin{figure*}[htb]
\centering
\includegraphics[width=0.9\linewidth]{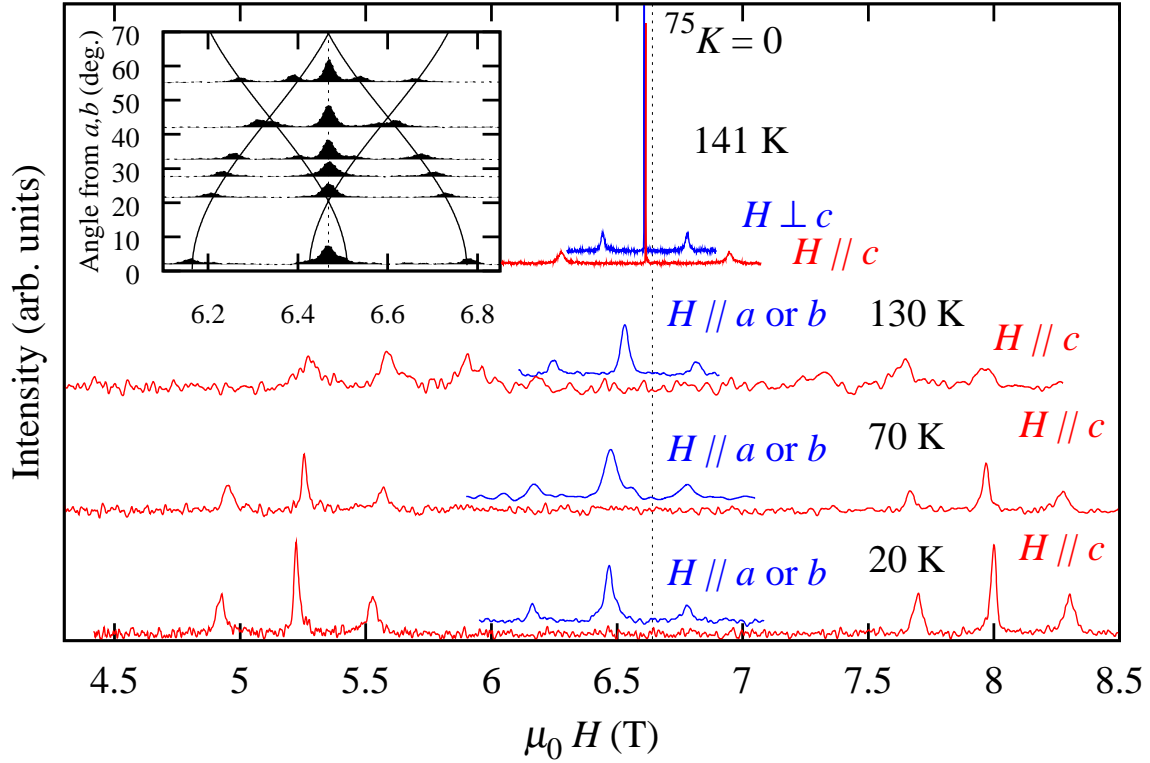}
\caption{(Color online) $^{75}$As-NMR spectra obtained by sweeping the magnetic field at the 
fixed frequency of 48.41~MHz. In the paramagnetic state at 141~K, the spectrum consists
of a sharp central line and two broader satellite lines. Below 135~K, the internal fields along the 
$c$-axis due to commensurate AF order splits the NMR spectra for $H \parallel c$ (red lines) 
and shifts the spectra to lower fields for $H \parallel a$- or $b$-axis (blue lines). 
The inset shows angular variation of the NMR spectrum at 20~K for the field 
rotated in the $ab$-plane.  The two sets of satellite lines originate from the 
twinned structural domains in the orthorhombic phase.  The satellite positions are fitted 
to Eq.~\eqref{eq:res} with the quadrupole parameters: $\nu^c = 2.21$~MHz,
$|\nu^a - \nu^b|/|\nu^c| = 1.18$ (solid lines).}
\label{fig:spectra}
\end{figure*}
Prior to the synthesis of the self-flux grown crystals, we also grew crystals
with Sn-flux\cite{NiBaK122SnFlux}. However, the energy dispersive x-ray fluorescence (EDX) spectroscopy revealed
1.5\% content of Sn substitution and the transition temperature detected by resistivity 
measurement was reduced to $\sim 70$~K.  Also the NMR spectra were extremely broad.

\section{Experimental Results}
The main panel of Fig.~\ref{fig:spectra} shows $^{75}$As-NMR spectra at 
the fixed frequency of 48.41~MHz obtained by sweeping the magnetic field
parallel and perpendicular to the $c$-axis at several different temperatures. 
The spectra at $T$=141~K are representative of the paramagnetic phase, while
the spectra at the other temperatures belong to the AF ordered phase.  

As $^{75}$As nuclei have spin 3/2, the NMR frequencies $\nu$ are generally
expressed by a perturbation series when the magnetic Zeeman interaction dominates over 
the quadrupole interaction,  
\begin{align}\label{eq:res}
\nu_{m \leftrightarrow m-1}
&= \mu_0\,^{75}\gamma H_\mathrm{eff}\nonumber\\ 
&+ \frac{1}{2}\nu^{c}\left(m-\frac{1}{2}\right)\bigg(3\cos^2\theta -
1\nonumber\\
&\quad\quad+ \frac{\nu^a - \nu^b}{\nu^c}\sin^2\theta\cos 
2\phi\bigg)\nonumber\\ &+ (\text{2nd \ order \ correction}).
\end{align}
The first term is the Zeeman frequency, where $^{75}\gamma =7.29019$~MHz/T is the 
nuclear gyromagnetic ratio and $H_\mathrm{eff}$ is the effective field at the
As nuclei. In the paramagnetic state, the effective field is expressed as $H_\mathrm{eff}= (1 + K^i) H$, 
where $H$ is the external field and $K^i$ is the Knight shift along the $i$-axis. 
The second term represents the first order quadrupolar shift for the three nuclear transitions 
$I_{z}=m \leftrightarrow m-1$ ($m$ =3/2, 1/2, or -1/2).  The explicit expression for the 
second order correction can be found in a standard
textbook\cite{MetallicShifts}. The quadrupole splitting parameters
$\nu^i\,(i=a,b,c)$ are related to the electric field gradient (EFG) tensor
$V_{\alpha \beta}=\partial^{2}V/\partial r_{\alpha}\partial r_{\beta}$ at the As nuclei as
$\nu^{i} = eV_{ii}Q/2h$, where $Q$ is the nuclear quadrupole moment of $^{75}$As.  Here
the crystal axes are defined with respect to the orthorhombic ($Fmmm$) unit cell.
Note that the $mm2$ symmetry of the As sites 
in the orthorhombic phase guarantees that the crystalline $a$, $b$, and $c$ axes 
are the principal axes of EFG tensor. The polar angle $(\theta, \phi)$ specifies the 
direction of the magnetic field relative to the crystalline axes. In the tetragonal phase, 
the NMR spectrum is independent of $\phi$ since $\nu_{a}=\nu_{b}=-\nu_{c}/2$. 

\begin{figure}[tb]
\centering
\includegraphics[width=0.9\linewidth]{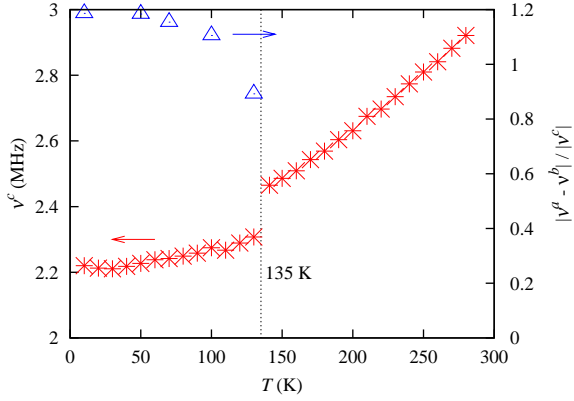}
\caption{(Color online) Nuclear quadrupole splitting frequency along the $c$ axis, $\nu^c$, and
the asymmetry parameter $|\nu^a - \nu^b| / |\nu^c|$ are plotted as a function of temperature.  
The asymmetry parameter is zero in the tetragonal phase. Its emergence below 135~K
and a jump in $\nu^c$ indicate the first-order structural transition into the orthorhombic phase.}
\label{fig:nuq}
\end{figure}

As shown in Fig.~\ref{fig:spectra}, the NMR spectra at $T$=141~K show a sharp central
line ($m$=1/2) with the full width at half maximum (FWHM) of 3.5~kHz and two broader
satellite lines ($m=3/2$ and $-1/2$) with the FWHM of 100~kHz.  We emphasize that 
the crystal grown in Sn-flux exhibits much broader lines (the FWHM is 150~kHz for the central line and
500~kHz for the satellite lines), indicating substantial disorder caused by
the substituted Sn. The quadrupole splitting $\nu_{c}$ is determined from the
separation between the two satellite lines for $H \perp c$ ($\theta=\pi/2$), which is a half of the separation for 
$H \parallel c$ ($\theta=0$) as expected. The value of $\nu_{c}$ is plotted against 
temperature in Fig.~\ref{fig:nuq}. 

\begin{figure}[tb]
\centering
\includegraphics[width=0.9\linewidth]{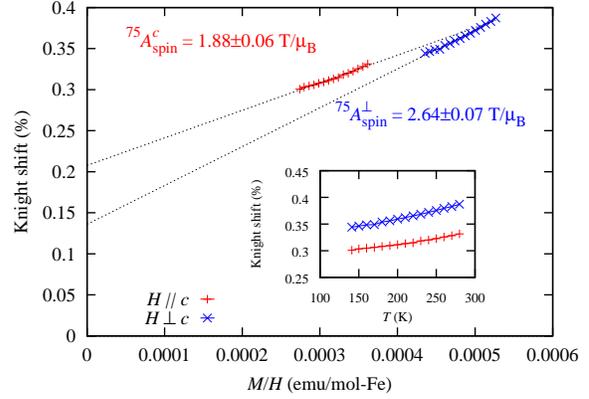}
\caption{(Color online) Inset: the temperature dependence of the Knight shift $^{75}K$ for 
the field parallel and perpendicular to the $c$-axis.  Main panel: $^{75}K$ 
is plotted against the bulk susceptibility $\chi$ measured by Wang
~\textit{et\,al.}\cite{WangBa122SC} The dotted lines represent the fits to a linear relation.}
\label{fig:kchi}
\end{figure}

The Knight shift $^{75}K$ was determined from the resonance position of the central line 
after correcting for the demagnetization field and the second order quadrupolar shift.   
The temperature dependence of $^{75}K$ is shown in the inset of Fig.~\ref{fig:kchi} for
$H \parallel c$ and for $H \perp c$ above 140~K. For both directions, $^{75}K$ decreases 
slightly with decreasing temperature.  This temperature dependence agrees with the earlier 
data on a crystal grown in Sn-flux by Baek~\textit{et\,al.}\cite{BaekNMR} and
on a powder sample by Fukazawa~\textit{et\,al.}\cite{FukazawaPolyAsNMR}
However, there are some discrepancies in the absolute values of the Knight shift.  In general, the Knight shift consists of the $T$-dependent 
spin shift, and the $T$-independent chemical (orbital) shift.  
\begin{equation}
K(T)=K_\mathrm{chem} + K_\mathrm{spin}(T) .
\end{equation}
Likewise the susceptibility is the sum of the contributions from core diamagnetism, orbital 
(van Vleck) paramagnetism, and spin paramagnetism,
\begin{equation}
\chi(T) = \chi_\mathrm{dia} + \chi_\mathrm{orb} + \chi_\mathrm{spin}(T) .
\end{equation}
The spin shift is linearly related to the spin susceptibility via the hyperfine coupling constant $A_\mathrm{spin}$,
\begin{equation}
K_\mathrm{spin}(T) =  A_\mathrm{spin} \chi_\mathrm{spin}(T)/ N_{A}\mu_\text{B}.
\end{equation}
We have indeed observed linear relations between $^{75}K$ and the magnetic susceptibility
as displayed in the main panel of  Fig.~\ref{fig:kchi}.  The slope of these
plots gives the values of the spin hyperfine coupling constant as $^{75}A^{ab}_\text{spin} =
2.64 \pm 0.07$~T/$\mu_\text{B}$, and $^{75}A^{c}_\text{spin} = 1.88 \pm 0.06$~T/$\mu_\text{B}$.

However, these values must be taken with caution.  Since $\chi_\mathrm{dia}\sim 
-6 \times 10^{-5}$~(emu/mol-Fe) and $\chi_\mathrm{orb}$ should be of the order of 
$1 \times 10^{-4}$~(emu/mol-Fe),  we expect $K_\mathrm{chem}$ to be larger
then the $y$-intercept of the K-$\chi$ plots in Fig.~\ref{fig:kchi}.  Although 
not much data are available for the chemical shift of As compounds,  the value 
$K_\mathrm{chem} \sim 0.2$~\% appears to be anomalously large. Therefore,
we cannot rule out the possibility that the hyperfine coupling is changing 
as a function of temperature.  If this is the case, the true values of
$A_\text{spin}(T)$ should be even larger than given above.
   
\begin{figure}[tb]
\centering
\includegraphics[width=0.9\linewidth]{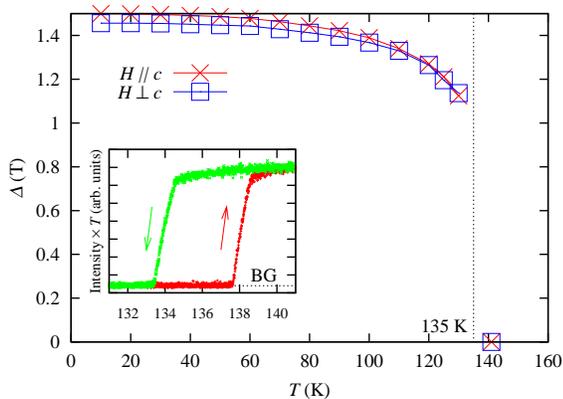}
\caption{(Color online) The temperature dependence of the internal field $\Delta$ 
at the As nuclei associated with the antiferromagnetic order. 
The inset shows the intensity of the central line at the resonance position 
of the paramagnetic phase for $H \perp c$.   The abrupt change 
with hysteresis is the evidence for the first-order magnetic transition.}
\label{fig:gap}
\end{figure}

When the temperature is lowered below 135~K, the NMR spectrum 
develops a two-fold splitting for $H \parallel c$, doubling the number of resonance lines 
(Fig.~\ref{fig:spectra}).  This is the direct evidence for a two-sublattice AF order,
consistent with the earlier M\"{o}ssbauer\cite{RotterBa122SDW}and neutron\cite{HuangBa122NS}
experiments. The AF moments generate a spontaneous internal field with alternating sign 
$\pm \Delta$, therefore, $H_\mathrm{eff}= H \pm \Delta$ in Eq.~\eqref{eq:res}.
The commensurate AF order has been reported previously by
Fukazawa~\textit{et\,al.} based on the $^{75}$As-NMR spectrum at zero magnetic field\cite{FukazawaPolyAsNMR}. 
The temperature dependence of $\Delta$ determined from the separation 
between the split center lines is shown in Fig.~\ref{fig:gap} by crosses. On the other hand, 
no splitting was observed when the field is applied along the $a$- or $b$-axis.
Instead the whole spectrum is shifted to lower fields (Fig.~\ref{fig:spectra}).  
This indicates that the internal field is parallel to the $c$-axis. When $H \perp c$, 
$H_\mathrm{eff}$ is given by the magnitude of the vector sum of the mutually
orthogonal external and internal fields, $H_\mathrm{eff}= \sqrt{H^{2} + \Delta^{2}}$, 
giving positively shifted unsplit resonance lines.  This allows us to determine $\Delta$ also
from the shift of the central line for $H \parallel a$ or $b$, as indicated by squares 
in Fig.~\ref{fig:gap}. The values of $\Delta$ for the two field orientations show good 
agreement, indicating that the magnitude and the direction of the AF moment are 
independent of the field direction.  

We have studied the behavior in the vicinity of the phase transition by
recording the peak intensity of the Fourier transformed spectrum at
the resonance position of the paramagnetic phase with changing temperature as shown in the inset of Fig.~\ref{fig:gap}.
The magnetic transition was found to be 
very sharp with the transition width less than 1~K, indicating good homogeneity of our crystal.  
It also shows a clear hysteresis in temperature about 4~K wide, providing conclusive 
evidence for the first-order transition. This is also inferred from the discontinuous 
development of the internal field.  
Let us remark that the crystal grown in Sn-flux shows much broader transition 
and extremely large line width (2~MHz) in the AF phase even though the magnitude 
of the internal field below 20~K is the same as observed for the self-flux grown crystal. 

We have also confirmed a discontinuous structural transition by a pronounced 
change of symmetry of EFG.  As shown in the inset of Fig.~\ref{fig:spectra}, 
the quadrupole splitting at 20~K varies with the field direction ($\phi$) in the $ab$-plane, 
indicating that $\nu^a \neq \nu^b$.  This is consistent with the orthorhombic symmetry. 
The angular variation of the NMR spectra displayed in the inset of Fig.~\ref{fig:spectra} shows 
two branches of satellite lines, which are shifted by 90$^{\circ}$.  This comes from the 
twinned orthorhombic domains.  The $\phi$ dependence of the satellite lines can be fit to 
Eq.~\eqref{eq:res}.  The extrema of the angular variation correspond to the 
field directions parallel to either $a$- or $b$-axis.  Note, however, that we cannot determine which
branch corresponds to the $a$- or $b$-axis.   The asymmetry parameter of EFG defined by 
$|\nu^a - \nu^b| / |\nu^c|$ is plotted in Fig.~\ref{fig:nuq}.  The asymmetry parameter, which is 
zero in the tetragonal phase, develops abruptly below the transition temperature.  A jump in 
$\nu_{c}$ is also observed across the transition.  These results establish the simultaneous
structural and magnetic transition.  

What is most surprising is the large value of asymmetry parameter exceeding one in the low 
temperature phase.  This means that the principal axis for the largest EFG changes by 
90$^{\circ}$ across the transition. Since the orthorhombicity of the lattice constant 
($b/a$) is less than 1\%\cite{RotterBa122SDW}, such a drastic change of EFG must be 
caused by substantial change of the charge density distribution around the As sites. 
This suggests that the band structure or the nature of the Fe-As bonds becomes highly
anisotropic in the $ab$-plane. We also notice anomalously large temperature variation
of $\nu_{c}$ in the paramagnetic phase.  Nearly 20\% reduction of $\nu_{c}$ from 300~K
to 140~K is clearly beyond (and in the opposite direction from) what is expected from 
normal thermal lattice contraction, indicating large variation in the local charge distribution
     
\begin{figure}[tb]
\centering
\includegraphics[width=0.9\linewidth]{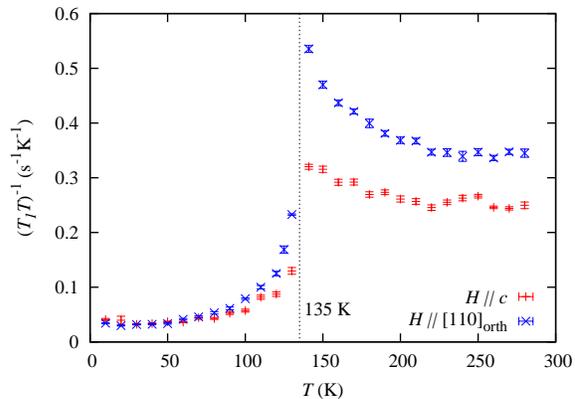}
\caption{(Color online) Nuclear spin-lattice relaxation rate divided by temperature, $(T_1T)^{-1}$, is 
plotted as a function of temperature for two field-orientations. The behavior, $T_1T = $const., 
at low temperatures indicates the itinerant (Fermi liquid) character of the material. We should remark that 
the upturn of $(T_1T)$ in the paramagnetic phase is observed only for $H \perp c$.  This indicates 
anisotropic development of the stripe-type antiferromagnetic spin fluctuations perpendicular to both 
the $c$-axis and the stripe.}
\label{fig:t1t}
\end{figure}
Finally, we show the spin-lattice relaxation rate divided by temperature $(T_1T)^{-1}$ in 
Fig.~\ref{fig:t1t}.  In the paramagnetic phase, $(T_1T)^{-1}$ shows anisotropic 
temperature dependence.  Substantial enhancement is seen for $H \parallel a$ or $b$ 
with decreasing temperature, while $(T_1T)^{-1}$ is nearly constant for 
$H \parallel c$.  Across the first order structural/magnetic transition, $(T_1T)^{-1}$ 
drops discontinuously by a factor of three.  At low temperatures, $(T_1T)^{-1}$ 
approaches a constant value for both field orientation expected for a Fermi liquid. 
This is consistent with the metallic behavior of the resistivity.  Thus the low temperature
phase is characterized as an itinerant antiferromagnetic phase.  Although 
the AF order may be driven by Fermi surface nesting, our results demonstrate that 
the order is commensurate and a sizable density of states at the Fermi level remains 
down to $T=0$.  

\section{Discussion}
Let us first discuss the nature of hyperfine interaction between As nuclei and Fe spins.
Generally this is the sum of the dipolar interaction and the transferred hyperfine interaction.
While the former is long ranged, the latter involves hybridization between the Fe-$3d$
and As-$4s$, $4p$ orbitals and it is usually sufficient to consider only the nearest neighbor 
interaction.  Since As-Fe bonds form nearly perfect tetrahedra, the dipolar field becomes 
extremely small, at most $0.01$~T/$\mu_\text{B}$ along the $c$-direction and a
half of this value perpendicular to the $c$-direction.  This is two orders of magnitude smaller than 
the experimental coupling constants in the paramagnetic state.  Therefore, dominant contribution 
comes from the transferred hyperfine fields.  

\begin{figure}[tb]
\centering
\includegraphics[width=0.9\linewidth]{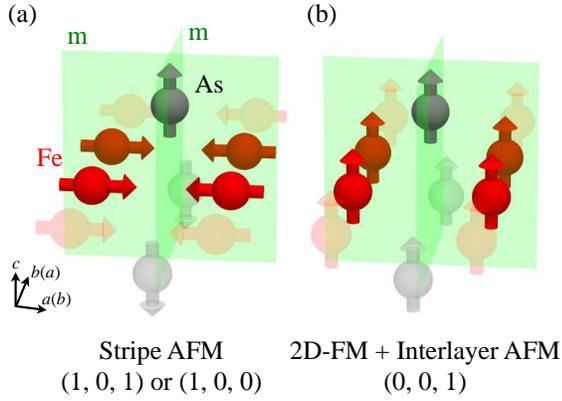}
\caption{(Color online) Possible spin configurations in the AF ordered state in
BaFe$_2$As$_2$, which is compatible with the $^{75}$As-NMR results. Only one
Fe-As layer is drawn for simplicity. The internal field at the As sites are shown by grey arrows. 
There are two candidates: (a) the stripe AF order and 
(b) AF stacking of the ferromagnetic layers. The former case 
(a) is realized in BaFe$_2$As$_2$ (see text).}
\label{fig:afpat}
\end{figure}
We now consider the spin structure in the AF ordered state which is compatible with
the $^{75}$As-NMR spectra.  There are two major experimental constraints: 
(i) the internal field at the As sites is parallel to the $c$-axis and 
(ii) The AF state has a simple two-sublattice structure. The formal symmetry 
analysis of the hyperfine field for various AF spin structure is presented in the Appendix,  
where we demonstrate that there are two possible spin structure compatible
with the NMR experiments as shown in Fig.~\ref{fig:afpat}.  More intuitively,
by considering the dipolar field from Fe spins, one can understand that only these two 
configurations generate a finite $c$-component of the internal field. In the first configuration  
(a), the AF order is described by the wave vector (101) or (100), forming stripes with the 
AF moments directed perpendicular to the stripe.  (Note that we cannot distinguish the 
orthorhombic $a$- and $b$-axes by NMR spectra.) In the second configuration (b), 
ferromagnetic layers in the $ab$-plane stack antiferromagnetically along the $c$-axis. 
The second structure, however, is highly unlikely since the temperature dependence of the
susceptibility rules out ferromagnetic correlation.

The above conclusion is fully consistent with the results of neutron diffraction 
experiment on BaFe$_2$As$_2$\cite{HuangBa122NS},
SrFe$_2$As$_2$\cite{KanekoSr122NS}, and CaFe$_2$As$_2$\cite{GoldmanCa122NS}. 
The AF order is described by the wave vector (101) in the orthorhombic structure.
The AF moment is found to be parallel to the $a$-axis. Note that the internal field 
along the $c$-axis at the As nuclei is generated only by the $a$-component of the 
AF moment (see the arguments in the Appendix). Therefore, only the anisotropic 
(off-diagonal) part of the hyperfine interaction can be relevant.  For example, the isotropic
Fermi contact interaction from the As-$4s$ orbital, which should be the dominant 
source of the transferred hyperfine interaction in the paramagnetic phase, does not 
contribute to the internal field in the AF ordered phase. A possible origin for such 
anisotropic hyperfine field is the dipolar field from the AF moments, which is calculated 
to be 0.25~T/$\mu_\text{B}$ for the stripe spin structure.  
Concerning the magnitude of the ordered moment, it is estimated
to be 0.89~$\mu_{B}$ from the neutron experiments\cite{HuangBa122NS}, while much smaller 
value 0.4~$\mu_{B}$ was proposed from the M\"{o}ssbauer experiments\cite{RotterBa122SDW}.
Even if we take the larger value of 0.9~$\mu_{B}$, the dipolar field from the AF
moments is nearly an order of magnitude smaller than the experimental value. Therefore,
the anisotropic hyperfine field from the AF moment must come from the hybridization
between the Fe-$3d$ and the As-$4p$ orbitals.  We should also recall that highly 
anisotropic charge density distribution associated with the Fe-As bonds is detected 
by the asymmetric EFG in the orthorhombic phase.  We suspect such anisotropic 
nature of the Fe-As hybridization in the orthorhombic phase may be the key 
ingredient to stabilize the stripe AF order.  

Motivated by possible coupling between AF order and structural distortion, 
we examined if the strong magnetic field affects the formation of twinned orthorhombic 
domains.  Since the AF moment, which is parallel to the orthorhombic $a$-direction, 
tends to align perpendicular to the external field, we expect that single domain 
structure may be achieved by field cooling.  However, this was not successful. 
We have cooled the crystal in the magnetic field of 11~T applied along [110] 
direction of the tetragonal unit cell from 140~K down to 40~K.  The angular 
variation of the NMR spectra at 40~K still showed two sets of satellite lines
from twinned domains with roughly equal intensity.   

We next discuss the anisotropic fluctuations observed in the relaxation rate measurements. 
Generally, $(T_1T)^{-1}$ can be expressed by the wave-vector-dependent 
dynamic spin susceptibility $\text{Im} \chi^{\perp} (q, \omega)$ and the hyperfine form factor $A_\perp(q)$ as 
\begin{equation}
\frac{1}{T_1T} \propto \lim_{\omega \rightarrow 0}\sum_{q}A^2_\perp(q)
\frac{\text{Im} \chi^\perp(q, \omega)}{\omega}.
\end{equation}
Note that the relaxation rate, which is the transition probability between the  nuclear spin
levels, is determined by the component of the dynamic susceptibility perpendicular to the 
effective field.  In the paramagnetic phase, enhancement of $(T_1T)^{-1}$ with decreasing 
temperature is observed only for $H \perp c$.  Therefore, this must be due to the 
fluctuations of the hyperfine field along the $c$-direction. As discussed in the appendix, 
such fluctuations can be generated only from the stripe-type spin fluctuations along the $a$-direction.
In other words, the fluctuations must be strongly anisotropic in the spin
space. Since the fluctuations are observed in the tetragonal paramagnetic phase, they might play an important role in the
superconductivity of the K-doped BaFe$_2$As$_2$, which does not exhibit
transition to the orthorhombic or antiferromagnetic phases.  If the stripe-type fluctuations exist in the superconducting materials, 
we expect the anisotropic enhancement of $(T_1T)^{-1}$ 
be observed in the $^{75}$As-NMR experiment. 

\section{Conclusions}
We have investigated the simultaneous magnetic and structural phase transition 
and magnetic fluctuations in the paramagnetic tetragonal phase 
in BaFe$_2$As$_2$ by $^{75}$As-NMR, using a high quality single crystal grown
by the self-flux method.  The spin structure in the AF ordered phase is determined 
almost uniquely to be of stripe type from the NMR spectra, in agreement with 
the neutron experiments.   

The substantial symmetry change of the electric field gradient (EFG) tensor at the As nuclei 
at the transition indicates drastic change of the charge distribution associated with 
the hybridization of Fe-$3d$ and As-$4p$ orbitals in spite of the apparently small
orthorhombic crystal distortion. The anisotropic nature of the Fe-As bonds may help
to stabilize the stripe AF order. The EFG shows strong temperature dependence even 
within the tetragonal phase, indicating continuous change of the local charge distribution.  

The NMR relaxation rates shows distinct behavior for different field orientations.
The enhancement of $(T_1T)^{-1}$ for $H \perp c$ with decreasing temperature
implies development of stripe antiferromagnetic fluctuations, which is anisotropic in spin space.
Since the stripe AF order appears by a first-order transition, persistence of such 
fluctuations in the paramagnetic phase is unexpected. Since the paramagnetic phase of 
BaFe$_2$As$_2$ evolves continuously to the superconducting phase by K doping,
 the stripe fluctuations may play an important role for superconductivity.  Whether this is the case
or not should be examined by the anisotropy of the $^{75}$As-NMR relaxation rates
in the superconducting materials in future studies.  

\section*{Acknowledgments}
We thank R.~Arita and Y.~Matsushita for enlightening discussions, and Y.~Kiuchi
for the EDX analysis. This work was supported partly by Grant-in-Aids on
Priority Areas ``Invention of Anomalous Quantum Materials'' (No. 16076204) 
and by Special Coordination Funds for Promoting 
Science and Technology “Promotion of Environmental Improvement for 
Independence of Young Researchers” from MEXT of Japan. 
K.\,K. and N.\,K. are financially supported as JSPS research fellows.

\appendix
\section{Symmetry Analysis of the Hyperfine Coupling Tensor and the Spin Structure in the AF phase}
\begin{figure}[tb]
\centering
\includegraphics[width=0.5\linewidth]{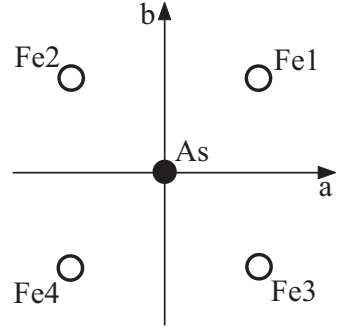}
\caption{Coordination of the nearest neighbor Fe sites around an As nucleus in
the orthorhombic unit cell. Note that As an Fe sites are not on the same plane.}
\label{fig:coupling}
\end{figure}

We discuss the spin structure in the AF phase based on the general symmetry
properties of  the hyperfine coupling tensor.  Since the long range dipolar interaction
gives only small contribution to the internal field, we focus on the short range transferred 
hyperfine interaction between the As nucleus and ordered moments on the four
nearest neighbor Fe sites (see Fig.~\ref{fig:coupling}) . 

The internal field can be written as the sum of contributions from 
each Fe sites.   
\begin{equation}
\bm{H}_\mathrm{int} =  \sum_{i=1}^{4} \bm{B}_{i}\cdot \bm{m}_{i} ,
\end{equation}
where $\bm{m}_{i}$ is the ordered moment at the $i$-th Fe site and $\bm{B}_{i}$ 
is the hyperfine coupling tensor between the As nucleus and $i$-th Fe site. 
We explicitly write components of $\bm{B}_{1}$ as  
\begin{equation}
    \bm{B}_{1}=  \left(
    \begin{array} {ccc}
    B_{aa} & B_{ab} & B_{ac} \\ 
    B_{ba} & B_{bb} & B_{bc} \\ 
    B_{ca} & B_{cb} & B_{cc}
\end{array}
\right) .
\end{equation}
Any component can be nonzero in the orthorhombic $Fmmm$ structure. 
The coupling tensor for other sites can be determined by symmetry consideration.  For example,
since the Fe2 site is related to the Fe1 site by mirror reflection with respect to the $bc$-plane,   
$\bm{B}_{2}$ is given as
\begin{equation}
    \bm{B}_{2}=  \left(
    \begin{array} {ccc}
    B_{aa} & -B_{ab} & -B_{ac} \\ 
    -B_{ba} & B_{bb} & B_{bc} \\ 
    -B_{ca} & B_{cb} & B_{cc}
\end{array}
\right) .
\end{equation}
Likewise 
\begin{eqnarray}
    \bm{B}_{3}=  \left(
    \begin{array} {ccc}
    B_{aa} & -B_{ab} & B_{ac} \\ 
    -B_{ba} & B_{bb} & -B_{bc} \\ 
    B_{ca} & -B_{cb} & B_{cc}
\end{array}
\right)  ,  \nonumber \\
    \bm{B}_{4} = \left(
  \begin{array} {ccc}
     B_{aa} & B_{ab} & -B_{ac} \\ 
    B_{ba} & B_{bb} & -B_{bc} \\ 
    -B_{ca} & -B_{cb} & B_{cc}
\end{array}
\right)  .
\end{eqnarray}
In the paramagnetic phase, the moment is uniform $\bm{m}_{i}=\bm{m}$,
therefore, $\bm{H}_\mathrm{int} = \left( \sum_{i=1}^{4} \bm{B}_{i}\right) \cdot \bm{m}$,
where 
\begin{equation}
\sum_{i=1}^{4} \bm{B}_{i}
=  4 \left( \begin{array} {ccc}
     B_{aa} & 0 & 0 \\ 
     0 & B_{bb} & 0 \\ 
    0 & 0 & B_{cc} 
\end{array}
\right)  .
\end{equation}
The diagonal components in the paramagnetic state can be determined from the 
experimental $K$ vs. $\chi$ plot as
$B_{aa} = B_{bb} = 0.66$~T/$\mu_{B}$, $B_{cc}=0.47$~T/$\mu_{B}$.  
 
Let us now examine the internal fields for all possible antiferromagnetic structure. 

\noindent \textit{(Case I)}: We first consider the stripe-order specified by the wave vector (100) or (101) 
shown in Fig.~\ref{fig:afpat}(a).  In this case, 
\begin{equation}
\bm{m}_{1} = -\bm{m}_{2} = \bm{m}_{3} = -\bm{m}_{4} \equiv \bm{\sigma}^\mathrm{I} ,
\end{equation}
therefore, 
\begin{equation}
\bm{H}_\mathrm{int} =  \left( \bm{B}_{1}-  \bm{B}_{2} + \bm{B}_{3}- \bm{B}_{4} \right) \cdot \bm{\sigma}^\mathrm{I} .
\end{equation}
Since 
\begin{equation}
\bm{B}_{1}-  \bm{B}_{2} + \bm{B}_{3}- \bm{B}_{4}
=  \left( \begin{array} {ccc}
     0 & 0 & 4B_{ac} \\ 
     0 & 0 & 0 \\ 
    4B_{ca} & 0 & 0
\end{array}
\right)  ,
\end{equation}
we obtain 
\begin{equation}
\bm{H}_\mathrm{int} =  4B_{ac} \left( \begin{array} {c}
\sigma^\mathrm{I}_{c} \\
0 \\
\sigma^\mathrm{I}_{a} 
\end{array} \right) .
\end{equation}
The internal field changes sign for the As site neighboring along the $a$-direction, 
resulting in the splitting of the NMR lines. In order to explain our NMR observation 
of the internal field parallel to the $c$-axis, the AF moment must be directed along 
the $a$-axis, which is perpendicular to the stripe direction.  This is entirely 
consistent with the results of neutron diffraction experiments\cite{HuangBa122NS}.
If we take the magnitude of the AF moment determined from the neutron experiments, 
$\sigma^\mathrm{I}_{a}= 0.87 \mu_{B}$ the experimental result $\Delta= 1.5$~T 
leads to the value $B_{ac}=0.43$~T/$\mu_{B}$.   

\noindent \textit{(Case II)}: We next consider the Neel order specified by the wave vector (110) or (111),
where $\bm{m}_{1} = -\bm{m}_{2} = -\bm{m}_{3} = \bm{m}_{4} \equiv \bm{\sigma}^\mathrm{II}$ . . 
Following the similar procedure, we have 
\begin{equation}
\bm{H}_\mathrm{int} =  \left( \bm{B}_{1}-  \bm{B}_{2} - \bm{B}_{3}+ \bm{B}_{4} \right) \cdot \bm{\sigma}^\mathrm{II}
\end{equation}
with  
\begin{equation}
\bm{B}_{1}-  \bm{B}_{2} - \bm{B}_{3}+ \bm{B}_{4}
=  \left( \begin{array} {ccc}
     0 & 4B_{ac} & 0 \\ 
     4B_{ac} & 0 & 0 \\ 
    0 & 0 & 0
\end{array}
\right)  . 
\end{equation}
We then obtain 
\begin{equation}
\bm{H}_\mathrm{int} =  4B_{ab} \left( \begin{array} {c}
\sigma^\mathrm{II}_{b} \\
\sigma^\mathrm{II}_{a} \\
0 
\end{array} \right) .
\end{equation}
Since the $c$-component of the internal field is zero for this case, it is not compatible with the NMR results.  

\noindent \textit{(Case III)}: For completeness, we consider the ferromagnetic moment 
in the $ab$-plane specified by the wave vector (001) as displayed in Fig.~\ref{fig:afpat}(b). 
As far as the single layer is concerned, this case is identical to the paramagnetic state already
considered.  The internal field is then given be
\begin{equation}
\Delta = 4B_{cc}\sigma^\mathrm{III}_{c} ,
\end{equation}
where $\sigma^\mathrm{III}_{c}$ is the $c$-component of the AF moment for this case.  The  
observed value $\Delta= 1.5$~T would then imply  $\sigma^\mathrm{III}_{c}= 0.80
\mu_{B}$.

\bibliography{document}

\end{document}